\documentstyle[prl,aps]{revtex}
\draft
\begin{document}
\input{epsf}
\twocolumn[\hsize\textwidth\columnwidth\hsize\csname@twocolumnfalse\endcsname
\title{Synchronous Behavior of Two Coupled Biological Neurons}
\author{Robert C. Elson$^{1,2}$, Allen I. Selverston$^{1,2,3}$,
Ramon Huerta$^{2,4}$, Nikolai F.Rulkov$^2$,\\
 Mikhail I. Rabinovich$^2$, and Henry D.I Abarbanel$^{2,5}$}
\address{
 $^1$ Department of Biology, University of California, San Diego, La
Jolla, CA 92093-0357\\
 $^2$Institute for Nonlinear Science, University of
California, San Diego, La Jolla, CA 92093-0402\\
 $^3$ Instituto de Neurobiologica, Old San Juan, 00901, Puerto Rico\\
 $^4$ ETS de Informatica, Universidad Aut\'{o}noma de Madrid,
 28409 Madrid, Spain \\
 $^5$ Department of Physics and Marine Physical Laboratory,
 Scripps Institution of Oceanography.\\ University of California,
 San Diego, La Jolla, CA 92093}

\date{\today}
\maketitle
\begin{abstract}
We report experimental studies of synchronization phenomena in a
pair of biological neurons that interact through naturally
occurring, electrical coupling. When these neurons generate
irregular bursts of spikes, the natural coupling synchronizes slow
oscillations of membrane potential, but not the fast spikes. By
adding artificial electrical coupling we studied transitions
between synchrony and asynchrony in both slow oscillations and fast
spikes. We discuss the dynamics of bursting and synchronization in
living neurons with distributed functional morphology.
\end{abstract}

\narrowtext
\vskip1pc]

The dynamics of many neural ensembles such as central pattern
generators (CPGs) or thalamo-cortical circuits pose questions
related to cooperative behavior of neurons. Individual neurons may
show irregular behavior~\cite{Glass95_Rabinovich98}, while
ensembles of different neurons can synchronize in order to process
biological information~\cite{Gray94_Meister91} or to produce
regular, rhythmical activity~\cite{harris92}. How do the irregular
neurons synchronize?  How do they inhibit noise and intrinsic
fluctuations? What parameters of the ensemble are responsible for
such synchronization and regularization? Answers to these and
similar questions may be found through experiments that enable one
to follow qualitatively the cooperative dynamics of neurons as
intrinsic and synaptic parameters are varied. Despite their
interest, these problems have not received extensive study. Results
of such an experiment for a minimal ensemble of two coupled, living
neurons are reported in this communication.

\begin{figure}
\begin{center}
\leavevmode
\hbox{%
\epsfxsize=5.7cm
\epsffile{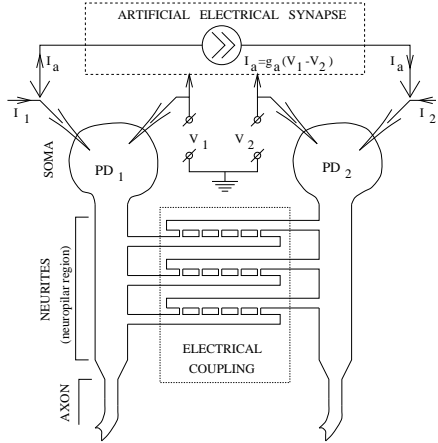}}
\end{center}
\caption{Schematic diagram of two coupled PD neurons. The stomatogastric ganglion (STG) of the
California spiny lobster,
{\em Panulirus interruptus} was removed using standard procedures
and pinned out in a dish lined with silicone elastomer and filled
with normal lobster saline~\protect\cite{foot1}. The STG remained
connected to its associated anterior ganglia, which provide
activating inputs~\protect\cite{harris92}. Separate, glass
microelectrodes (filled with 3M-KCI; tip resistance 10-20
M$\Omega$) were  inserted in the soma of each neuron, for
intracellular voltage recording or current injection. Each
microelectrode was served by a separate amplifier and current
source. Measured voltage signals were digitized at 5000
samples/sec. The two PD neurons remained coupled to each other by
their natural electrical synapses, but were isolated from the rest
of the CPG by blocking chemical input synapses with picrotoxin
($7.5\mu $M) and photo-inactivating other, electrically-coupled
neurons~\protect\cite{foot2}. Artificial electrical coupling was
provided by injecting equal and opposite current $I_a$ into the two
neurons, such that $I_a^{(j)}=g_a(V_j-V_i)$, where $g_a$ is the
added synaptic conductance and $V_i$ is the membrane potential at
the soma of PD$_i$~\protect\cite{foot3}.
\label{fig1}}
\end{figure}

The experiment was carried out on two electrically coupled neurons
(the pyloric dilators, PD) from the pyloric CPG of the lobster
stomatogastric ganglion~\cite{harris92}. Individually, these
neurons can generate spiking-bursting activity that is irregular
and seemingly chaotic. This activity pattern can be altered by
injecting DC current ($I_1$~and~$I_2$) into the neurons, see
Fig.~\ref{fig1}. In parallel to their natural coupling, we added
artificial coupling by a dynamic current clamp
device~\cite{sharp}. Varying these control parameters (offset
current and artificial coupling), we found the following regimes of
cooperative behavior.

\begin{figure}
\begin{center}
\leavevmode
\hbox{%
\epsfxsize=7.0cm
\epsffile{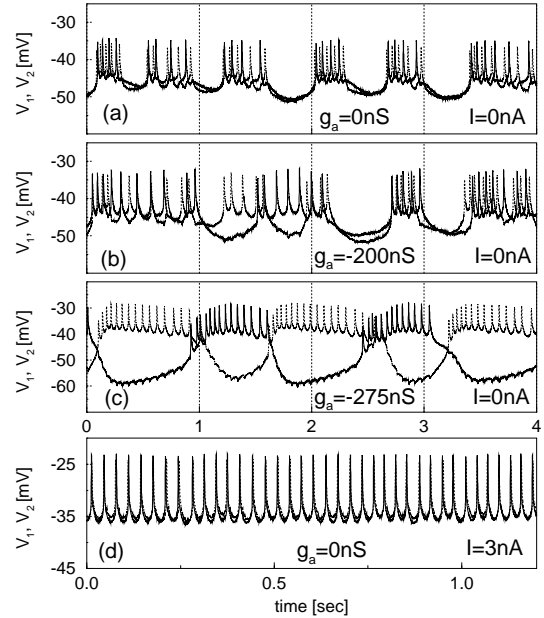}}
\end{center}
\caption{Regimes of oscillations in two coupled neurons.
\label{fig2}}
\end{figure}

Natural coupling produces state-dependent synchronization, see
Fig.~\ref{fig2}. $(i)$ When depolarized by positive DC current,
both neurons fire a continuous pattern of synchronized spikes
(Fig.~\ref{fig2}d). $(ii)$ With little or no applied current, the
neurons fire spikes in irregular bursts: now the slow oscillations
are well synchronized while spikes are not (Fig.~\ref{fig2}a).
Changing the magnitude and sign of electrical coupling restructures
the cooperative dynamics. $(iii)$ Increasing the strength of
coupling produces complete synchronization of both irregular slow
oscillations and fast spikes (see below). $(iv)$ Compensating the
natural coupling leads to the onset of independent irregular
pulsations (Fig.~\ref{fig2}b). $(v)$ With net negative coupling,
the neurons burst in antiphase, in a regularized pattern
(Fig.~\ref{fig2}c).

Figure~\ref{fig1} summarizes the functional geometry of this
pair~\cite{hartline92}. Each PD cell is a motor neuron, consisting
of a soma, a primary neurite and a neuropilar region, and an axon
which conducts spikes to target muscles. Within this extended
structure there is (1) frequency-dependent filtering of voltage
signals; and (2) spatial localization of active membrane
currents~\cite{hartline92}. First, the neurites constitute a cable.
When coupled to its partner, each neuron shows an input resistance
and capacitance of order 5M$\Omega$ and 5$\mu$F,
respectively~\cite{hartline92}. Second, the fast sodium and
potassium channels underlying threshold-dependent spike generation
(action potentials: amplitude $\approx100$mV; duration
$\approx1$msec) are concentrated in the membrane of the axon,
whereas the sodium, calcium and potassium channels supporting slow
voltage oscillations (10-30mV, 0.3-1.0sec) are located in the
neuropil~\cite{hartline92}. Cable properties affect the passive
spread of voltage signals within the neuron. Slow voltage
oscillations experience little attenuation in spreading from the
neuropil to the axon, where they drive bursts of spikes. However,
fast spike potentials suffer significant low-pass filtering as they
spread from the axon to the neuropil and soma~\cite{hartline92}.
The neuropil is also the site of natural electrical coupling
(non-rectifying and moderately weak: steady-state voltage
attenuation $\approx0.25$)~\cite{foot4}.

In the studies of synchronization in the coupled PD neurons both
cells were active under symmetrical DC current injection
($I_1=I_2=I$). Long records of $V_1(t)$ and $V_2(t)$ were obtained,
from which we show segments (Fig.~\ref{fig2}). For different values
of the parameter $I$ we see a regime of bursting-spiking
($I=0\,$nA, Fig.~\ref{fig2}a-c), and, at more depolarized levels, a
region of pure spiking activity ($I= 3\,$nA, Fig.~\ref{fig2}d).

\begin{figure}
\begin{center}
\leavevmode
\hbox{%
\epsfxsize=7.0cm
\epsffile{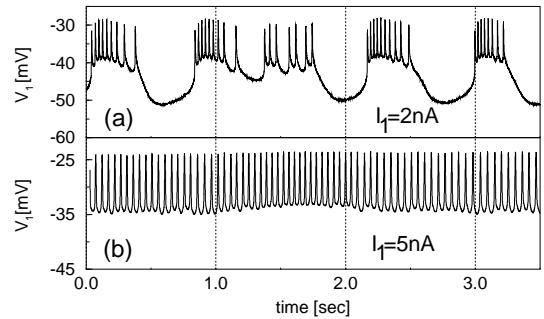}}
\end{center}
\caption{Time series for an isolated PD for different
values of $I_1$. $PD_2$ is inactivated by large negative current
injection.
\label{fig3}}
\end{figure}

Our results indicated that spiking and bursting--spiking regimes of
activity arise from the autonomous dynamics of individual PD
neurons. In the experiment (Fig.~\ref{fig3}a, b) we recorded the
membrane potential of one PD cell $PD_1$ when its partner $PD_2$
was deactivated by DC hyperpolarization to $-80$mV, effectively
suppressing its neural activity. The DC current $I_1$ injected into
$PD_1$ was varied. At $I_1 = 2\,$nA the activity consisted of
aperiodic slow oscillations surmounted by spikes
(bursting-spiking), see Fig.~\ref{fig3}a. At $I_1=5\,$nA, the
neuron generated fast spikes alone, see Fig.~\ref{fig3}b. Thus, the
voltage-dependent spiking and bursting properties of a single PD
resembled those of the active pair (although the values of $I_1$
are shifted relative to those of a symmetrical pair, due to the
shunting action of the deactivated neuron) (Fig.~\ref{fig2}a,d).
Similar, voltage-dependent activity regimes were also seen after
isolating a single PD neuron by photo-inactivating its partner.

\begin{figure}
\begin{center}
\leavevmode
\hbox{%
\epsfxsize=3.6cm
\epsffile{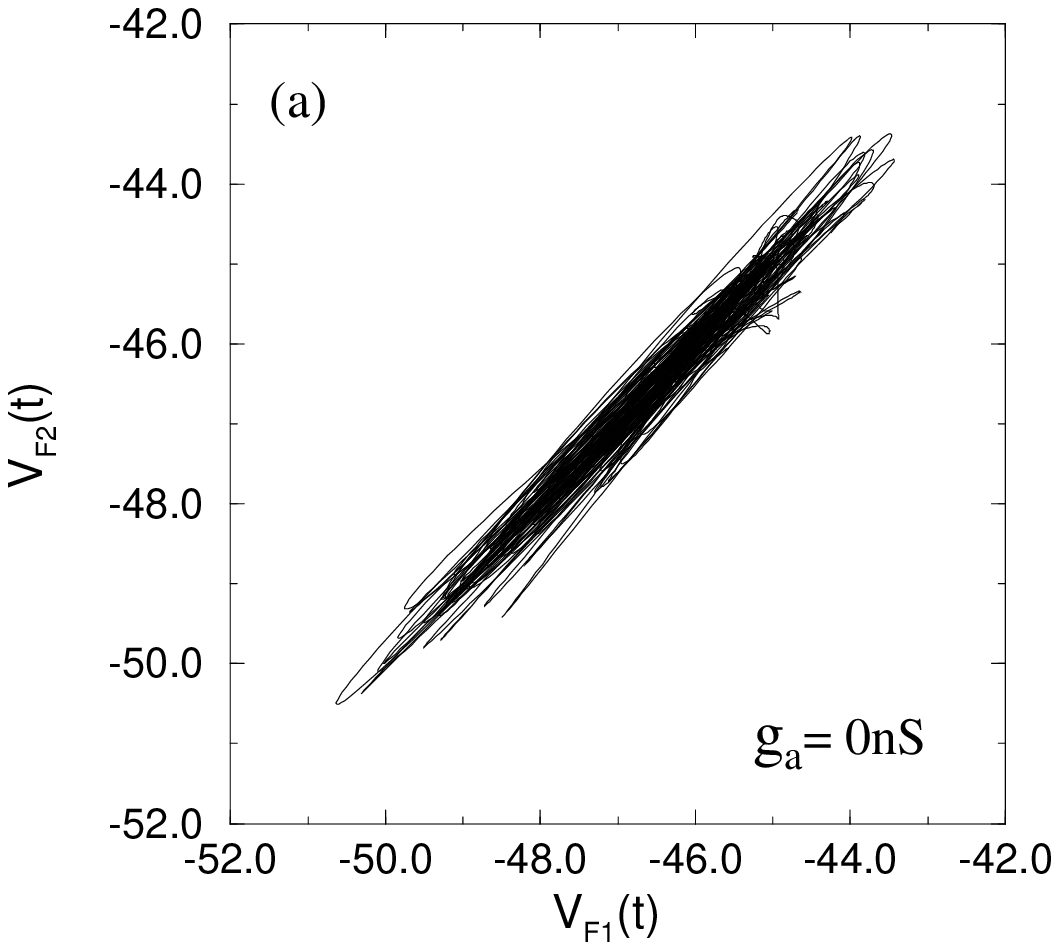}
\epsfxsize=3.6cm
\epsffile{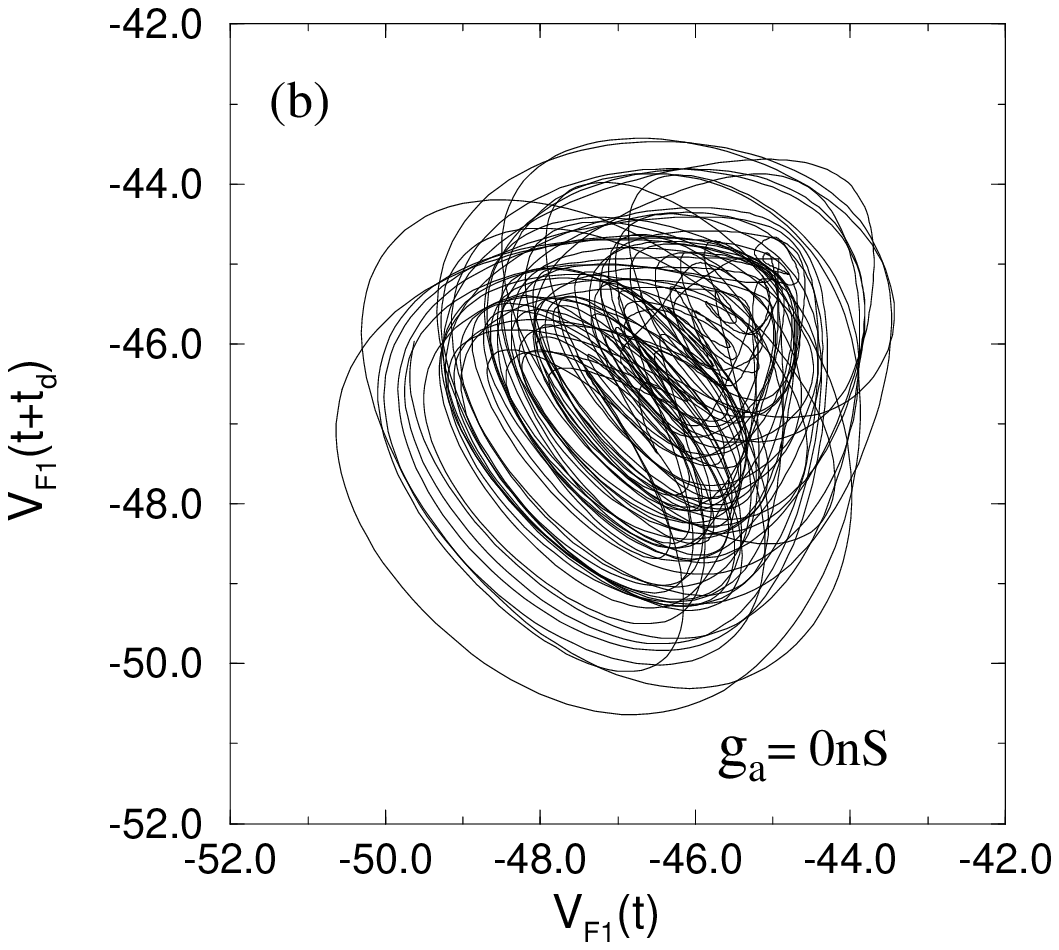}}
\hbox{%
\epsfxsize=3.6cm
\epsffile{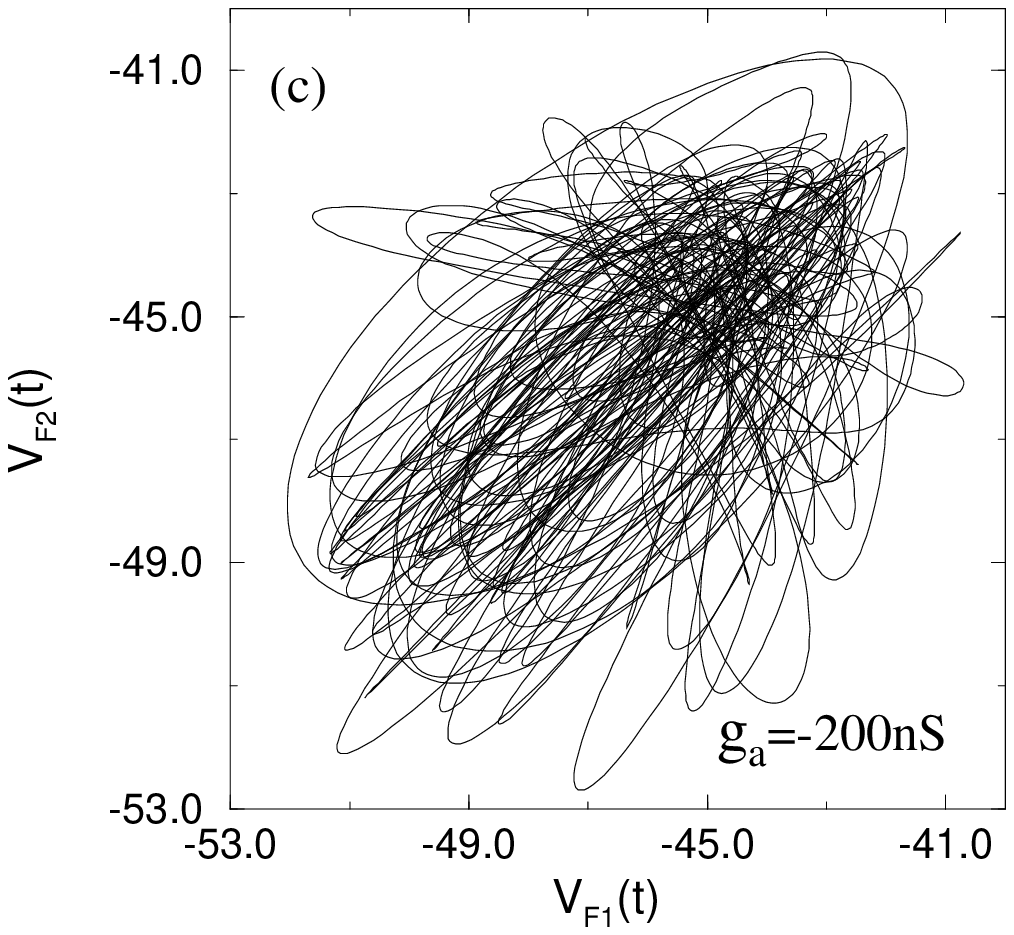}
\epsfxsize=3.6cm
\epsffile{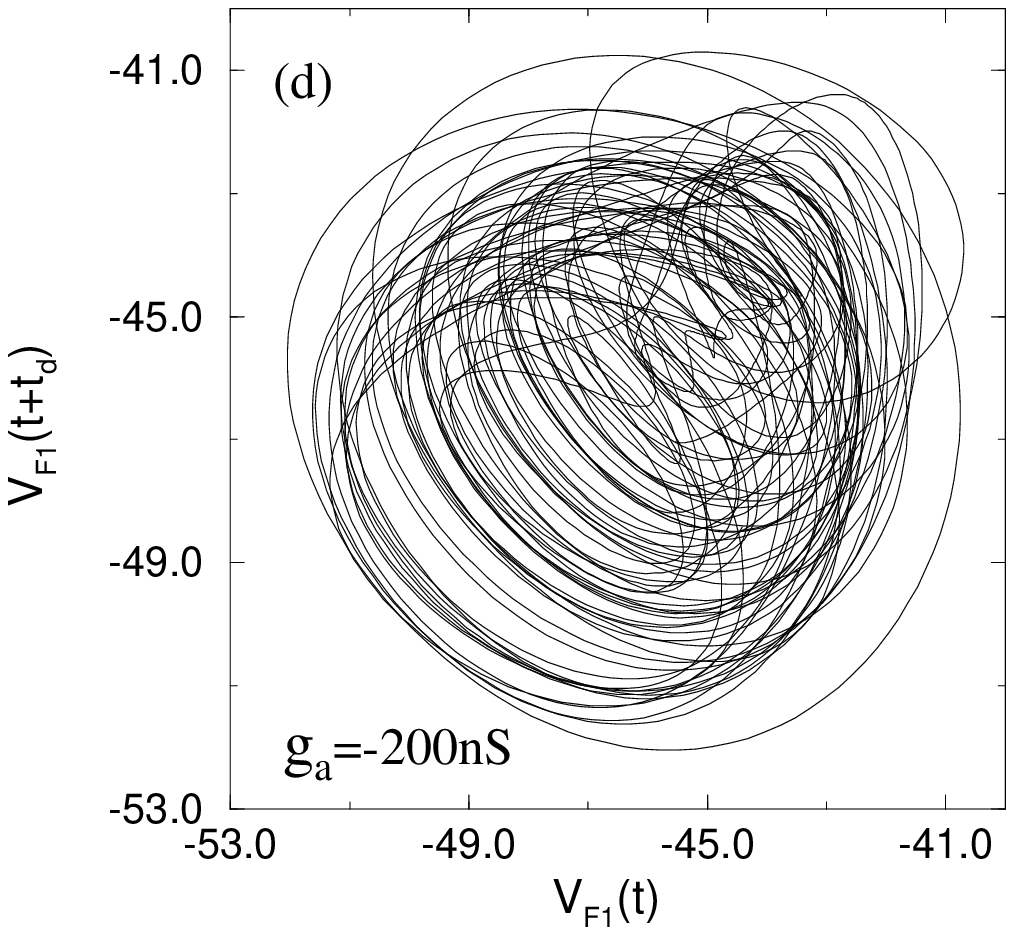}}
\hbox{%
\epsfxsize=3.6cm
\epsffile{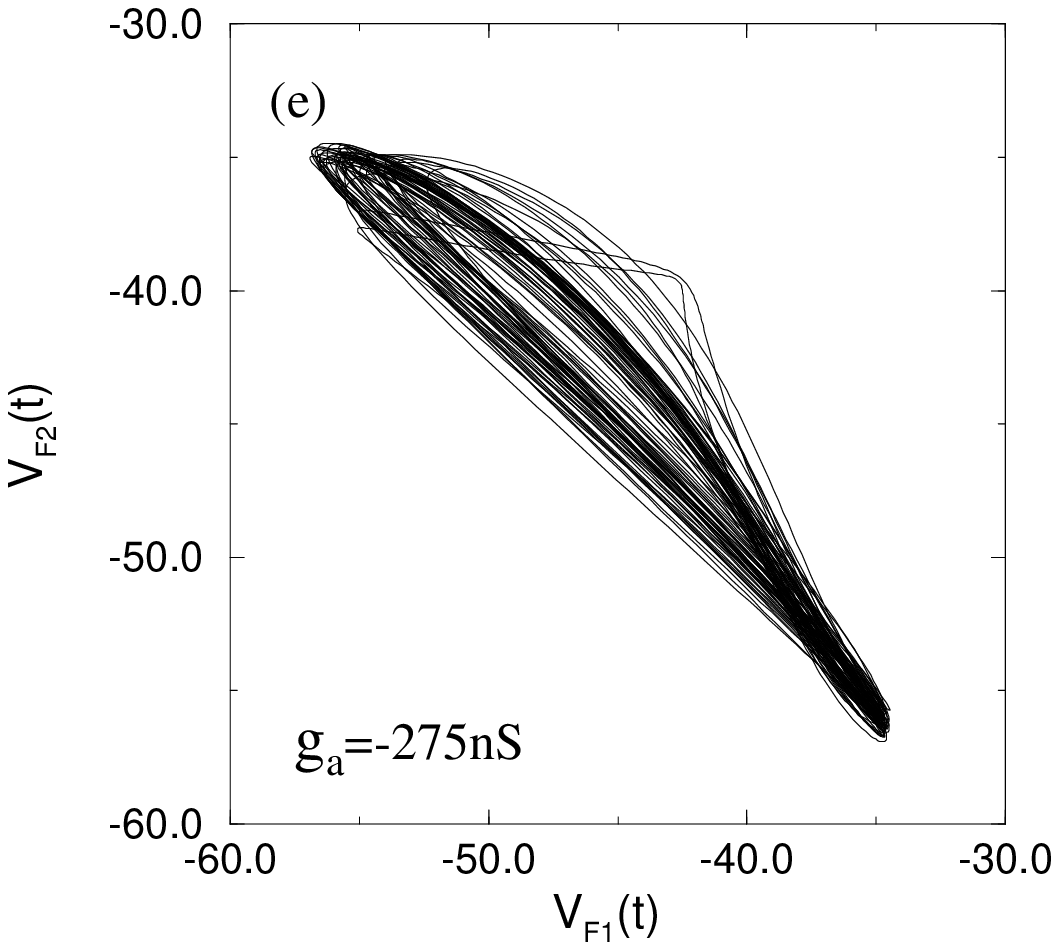}
\epsfxsize=3.6cm
\epsffile{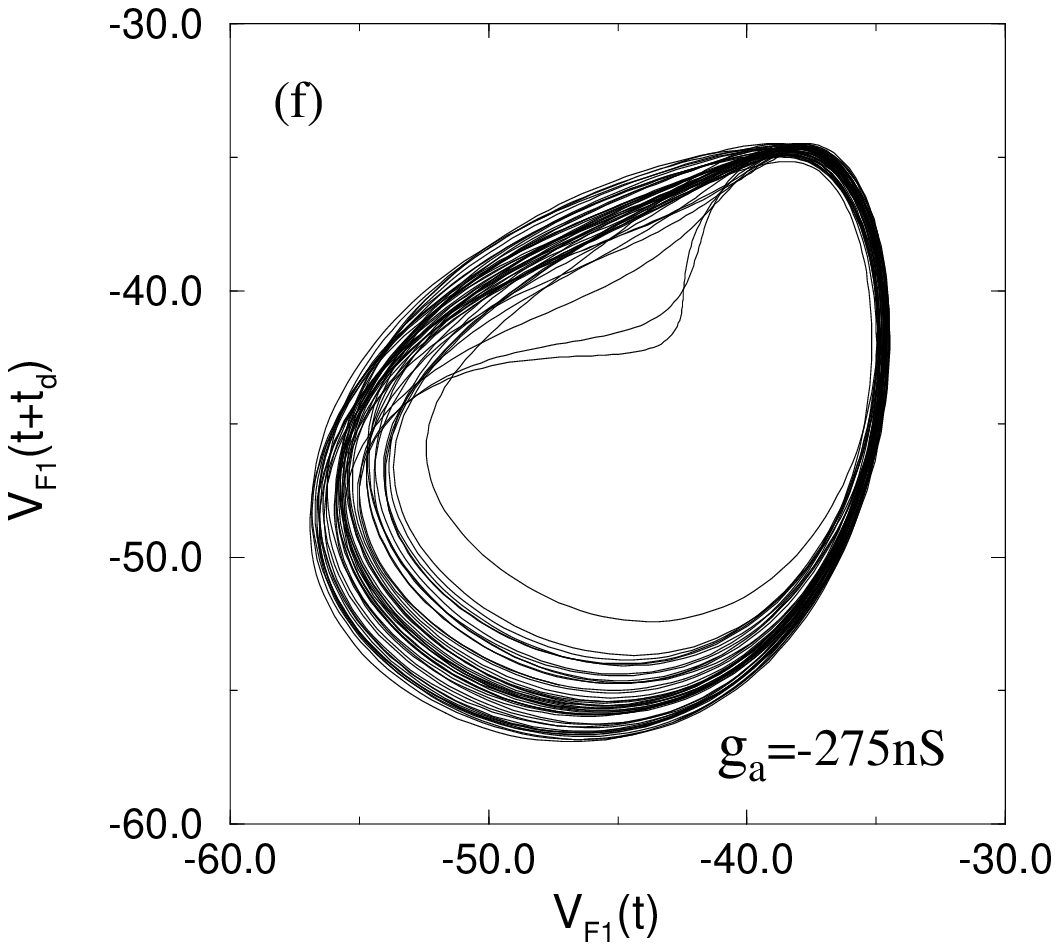}}
\end{center}
\caption{Phase portraits of the slow components of oscillations in
coupled neurons plotted in the planes variables
$(V_{F1}(t),V_{F2}(t))$--left, and
$(V_{F1}(t),V_{F1}(t+t_d))$--right. $t_d=0.3$sec. $I$=0
\label{fig4}}
\end{figure}

\begin{figure}
\begin{center}
\leavevmode
\hbox{%
\epsfxsize=7cm
\epsffile{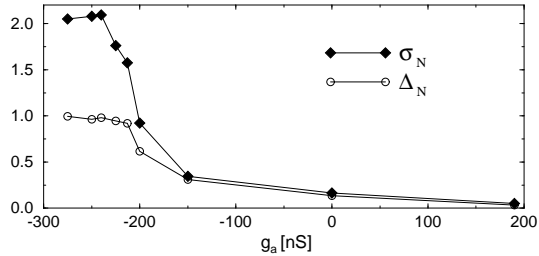}}
\end{center}
\caption{$\sigma_N$ and $\Delta_N$ as a function of the conductivity
through the electrical clamp $g_a$.
\label{fig5}}
\end{figure}

The records presented in Fig.~\ref{fig2}a clearly indicate the
synchrony of bursts in the naturally coupled neurons. For more
detailed analysis of synchronization of these aperiodic bursts we
adopt the technique developed for the experimental studies of chaos
synchronization in electronic circuits~\cite{circ}. To study the
synchronization of slow bursts we suppress the spikes in the
recorded signals using low--pass filter with cut--off frequency 5Hz
and analyze the "slow trajectories" given by the filtered signals
$V_{F1}(t)$ and $V_{F2}(t)$. The projections on to the planes of
variables ($V_{F1}(t),V_{F2}(t)$) and ($V_{F1}(t),V_{F1}(t+t_d)$)
shown in Fig.~\ref{fig4} characterize the level of synchrony of
bursts in the neurons and the complexity of the bursts dynamics,
respectively. To quantify synchronization, we calculate the
difference $V_{FD}(t)=V_{F1}(t)-V_{F2}(t)$, and study the
normalized standard deviation
$\sigma_N=\sigma_{V_{FD}}/\sigma_{V_{F1}}$ and normalized maximal
deviation $\Delta_N=|V_{FD}|^{max}/(V_{F1}^{max}-V_{F1}^{min})$ as
a function of $g_a$, see Fig.~\ref{fig5}.

The dynamics of slow oscillations changed as the effective coupling
conductance was altered by adding artificial coupling, $g_a$. With
natural coupling ($g_a=0$nS) the slow oscillations stayed
synchronized (Fig.~\ref{fig4}a) despite very complex dynamics
(Fig.~\ref{fig4}b) (cf. Fig.~\ref{fig2}a). Additional dissipative
coupling ($g_a>0$nS) increased the level of synchrony between the
neurons, while compensation of natural coupling ($g_a<0$nS) led to
desynchronization (Fig.~\ref{fig5}). The desynchronized, slow
oscillations remained complex and aperiodic (Fig.~\ref{fig4}c,d,
see also Figs.~\ref{fig2}b). Adding further, negative coupling
conductance ($g_a<-240$nS: probably overcompensating the natural
synapse) caused the neurons to become synchronized again, but in
antiphase (Figs.~\ref{fig4}e~,~\ref{fig5}). This regime of
antiphase synchronization was characterized by the onset of more
regular,``almost periodic" bursts (Fig.~\ref{fig4}f).

\begin{figure}
\begin{center}
\leavevmode
\hbox{%
\epsfxsize=8cm
\epsffile{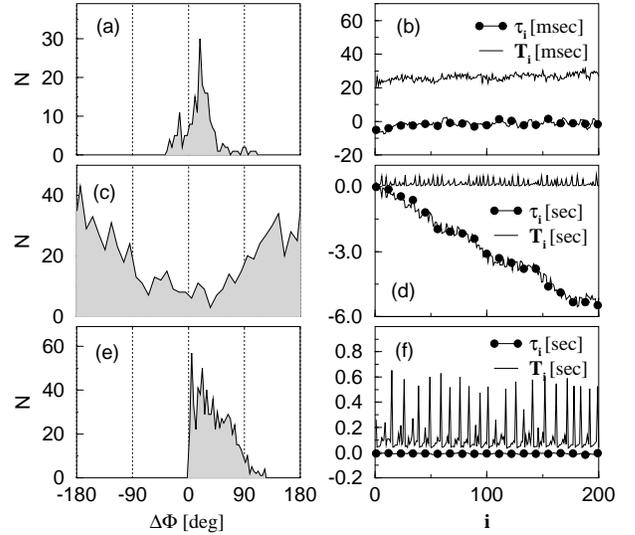}}
\end{center}
\caption{Analysis of synchrony in the fast spiking activity
measured for three different regimes of neuron activity described in the text.
Distributions of the phase lags between the spiking of the neurons
(left). The evolution of interspike intervals $\tau_i$ and $T_i$
(right).
\label{fig6}}
\end{figure}

Next we describe the synchronization of the fast, spike
oscillations. The standard criterion for identical synchronization
fails here because of small fluctuations in spike timing.
Therefore, we applied a different analysis. For each membrane
voltage we located the times of spike peaks, $t^{(1)}_i$ for $PD_1$
and $t^{(2)}_i$ for $PD_2$, and calculated the intervals
$T_i=t^{(1)}_{i+1}-t^{(1)}_i$ and $\tau_i=t^{(2)}_i-t^{(1)}_i$.
$T_i$ measures interspike intervals in $PD_1$, while $\tau_i$ tells
us about the difference in spike timing in $PD_1$ versus $PD_2$. If
$|\tau_i|$ did not grow with time and $\mbox{max}\{|\tau_i|\} <
\mbox{min}\{T_i\}$, we concluded that the neurons spike
synchronously. We also analyzed the level of synchronization by
measuring the phase relation between spikes. In this analysis we
plotted a histogram of the phase, $\Delta\Phi_i$, of the $i{th}$
spike of $PD_2$ within the interval formed by the neighboring pair
of spikes in $PD_1$ (designated $j{th}$ and $k{th}$, respectively),
using the function,
$\Delta\Phi_i=180^o(t^{(2)}_i-t^{(1)}_j)/|t^{(1)}_j-t^{(1)}_k|$.
The results of these analyses are shown in Fig.~\ref{fig6}.

With natural coupling, spikes were synchronized during tonic
firing. Figure~\ref{fig6}a~and~b shows such a synchronous regime
($I = 3$nA and $g_a=0$) where all values of $\Delta\Phi$ are within
the interval $(-40^o,130^o)$, and the value of $\tau_i$ oscillates
but remains smaller than $T_i$, see Fig.~\ref{fig2}d. In the
spiking-bursting regime, natural coupling did not synchronize the
spikes (unlike the slow oscillations) (Figs.~\ref{fig6}c~and~d). In
this condition ($I=0$nA and $g_a=0$: cf. Fig.~\ref{fig2}a), the
constant drift of $\tau_i$ indicated a difference in spike
frequencies; nevertheless, the non-uniform distribution of
$\Delta\Phi_i$ indicated that the neurons were not far from the
threshold for spike synchronization. Indeed, the spikes of bursting
neurons became synchronized when artificial (positive) coupling was
added (Figs.~\ref{fig6}e~and~f: $I=0$ and $g_a=190$nS).

Our observations indicate that the slow oscillations and fast
spikes of these two neurons have different thresholds for the onset
of synchronization. This can be understood in terms of the
different sites of origin of the two types of voltage signal, the
different mechanisms of synchronization, and the different
conduction pathways and attenuation factors involved (cf.
Fig.~\ref{fig1} and associated text). The slow voltage oscillations
that underlie bursting activity arise as a result of
voltage-dependent ion channel activity in the membrane of
neuropilar processes. The summed voltage signal will suffer some
attenuation as it spreads by local current flow in the leaky cable
array of the neuropil. However, two factors favor its effective
transmission between the neurons: the location of electrical
coupling sites close to the site of slow wave generation, and the
slow timecourse of the voltage signal itself. In combination, these
should allow a relatively strong and continuous interaction between
the irregular slow oscillators. This mechanism resembles the
synchronization seen in dissipatively coupled chaotic electrical
circuits~\cite{circ}. In contrast, fast spike signals suffer strong
attenuation as they spread between the spike initiation zone at the
origin of the axon and the coupling sites in the neuropil. These
factors argue for weak current flow between spike generators. If
the spike generator of one neuron is close enough to its threshold,
the transient current from the coupling pathway may drive it to
phase-locked firing. In electrical circuits, this type of chaotic
pulse synchronization is known as threshold synchronization
~\cite{threshold}. With natural coupling, this threshold mechanism
can synchronize spike activity in tonic firing but not in the
bursting regime. When the neurons generate slow voltage
oscillations, ion channel open in neuropilar processes, decreasing
the membrane resistance: this shunts the spike-evoked currents as
they flow in their coupling pathway, causing a failure in threshold
synchronization.

As the strength of net coupling is decreased, the slow oscillations
remain irregular with little change in waveform, but make a sharp
transition from synchronous to asynchronous behavior, see
Figs~\ref{fig4},~\ref{fig5}. When the net coupling reaches an
expected, negative conductance, the slow oscillations
re-synchronize in antiphase and become regular. These bifurcations
argue for a dynamical origin of the irregular neuronal activity.
Based on these observations we have built a two-compartment model
of the stomatogastric neuron. The model incorporate six active
ionic currents distributed in soma-neuropil and axon, and takes
into account slow, intercellular Ca$^{++}$ dynamics. Two such model
neurons, when electrically coupled, reproduce all five types of
behavior found in our experiments and the transitions between the
regimes are consistent with the observations reported here, see
Ref.~\cite{abarbanel}.

This work was supported by NIH grant NS09322 and NSF grant
IBN-9122712 and U.S. Department of Energy grants DE-FG03-90ER14138
and DE-FG03-96ER14592.


\begin{thebibliography}{99}

\bibitem{Glass95_Rabinovich98} L. Glass, in{\em The Handbook of
Brain Theory and Neural Networks}, Ed. M. Arbib (MIT Press,
Cambridge, MA), 186 (1995); M.I. Rabinovich and H.D.I. Abarbanel
{\em Neuroscience} {\bf 87}, 5 (1998).

\bibitem{Gray94_Meister91} C.M. Gray {\em J. Comput. Neurosci.} {\bf 1},
11 (1994); M. Meister {\em Science} {\bf 252}, 939 (1991).

\bibitem{harris92} R. M. Harris-Warrick, {\em et al}, {\em Dynamic
Biological Networks: The Stomatogastric Nervous System}, (MIT
Press, Cambridge, MA), 1992.

\bibitem{sharp} A.A. Sharp, L.F. Abbott, and E. Marder, {\em J.
Neurophysiol.} {\bf 67}, 1691 (1993).

\bibitem{hartline92} D. K. Hartline and K. Graubard  In: {\em
Dynamic Biological Networks: The Stomatogastric Nervous System}
eds. R. M. Harris-Warrick et al., (1992) pp. 31 - 85; K. Graubard
and D.K. Hartline,{\em Brain Research} {\bf 557}, 241 (1991); D.H.
Edwards, Jr. and B. Mulloney,{\em J. Physiol.} {\bf 348}, 89
(1984).

\bibitem{foot4} Measurements of the magnitude of the natural
coupling conductance is complicated by axial and shunting (leak)
conductances that intervene between the sites of electrode
insertion (somata) and electrical coupling (fine neurites).
Estimates center around 100-200nS: cf. M. Gola and A. I.
Selverston, {\em J. Comp. Physiol.} {\bf 145}, 191 (1981); B.R.
Johnson, J.H. Peck and R.M. Harris-Warrick, {\em J. Comp. Physiol.}
{\bf 172} 715 (1993).

\bibitem{foot1} The STG received separate superfusion with
saline whose temperature variation was maintained within $\leq
1^o\,$C during recording sessions (ranges between experiments,
$14^o$ to $17^o\,$C). See for details B. Mulloney and A. I.
Selverston,{\em J. Comp. Physiol.} {\bf  91} , 1 (1974).

\bibitem{foot2} Details of the procedure for synaptic isolation are
found in T. Bal, F. Nagy, and M. Moulins, {\em J. Comp. Physiol.}
{\bf 163}, 715 (1988); J.P. Miller and A.I. Selverston {\em J.
Neurophysiol.} {\bf 48}, 1378 (1982).

\bibitem{foot3} The artificial synapse was implemented by an analog
circuit as in~\cite{sharp} or by a computer-based dynamic current clamp
as in A.A. Sharp, M.B. O'Neil, L.F. Abbott, and E. Marder,
{\em J. Neurophysiol.} {\bf 69}, 992 (1993).

\bibitem{circ} V.S. Afraimovich, N.N. Verichev and
M.I. Rabinovich {\em Radiophys. Quant. Electr.} {\bf 29}, 747
(1986); J.F. Heagy, L.M. Pecora and T.L. Carroll {\em Phys. Rev. E}
{\bf 50}, 1874 (1994); N.F. Rulkov {\em et al}, {\em Int. J. Bif.
and Chaos} {\bf 2}, 669 (1992)

\bibitem{threshold} N. F. Rulkov and A. R. Volkovskii {\em Physics
Letters A} {\bf 179}, 332 (1993); Carroll, T.L.,
{\em Biological Cybernetics}, {\bf 73}, 553 (1995).

\bibitem{abarbanel} M.I. Rabinovich {\em et al}, {\em Physica} A, (1998)
(to be published).


\end{thebibliography}
\end{document}